\documentclass[runningheads]{llncs}
\usepackage[T1]{fontenc}
\usepackage{enumitem}
\usepackage{multirow}
\usepackage{graphicx}
\usepackage{amsfonts}
\usepackage{booktabs}
\usepackage{subcaption}
\usepackage{caption}
\usepackage{amsmath}
\usepackage{amssymb}
\usepackage{mathtools}
\begin{document}

\makeatletter
\renewcommand\section{\@startsection{section}{1}{\z@}%
  {3ex \@plus -0.5ex \@minus -.2ex} 
  {1ex \@plus .2ex} 
  {\normalfont\large\bfseries}}
\makeatother

\setlength{\abovedisplayskip}{4pt}
\setlength{\belowdisplayskip}{4pt}
\setlength{\abovedisplayshortskip}{2pt}
\setlength{\belowdisplayshortskip}{2pt}

\title{FGR-ColBERT: Identifying Fine-Grained Relevance Tokens During Retrieval}

%
%

\author{
Antonín Jarolím\thanks{Correspondence to: \texttt{ijarolim@fit.vut.cz}\\Preprint. Work-in-progress.} and
Martin Fajčík  
}
\institute{Brno University of Technology, Czech Republic}
\authorrunning{A. Jarolím et al.}

%
\maketitle              
\vspace{-1em}
\begin{abstract}

Document retrieval identifies relevant documents but does not provide fine-grained evidence cues, such as specific relevant spans. A possible solution is to apply an LLM after retrieval; however, this introduces significant computational overhead and limits practical deployment.
We propose FGR-ColBERT, a modification of ColBERT~\cite{santhanam2022colbertv2} retrieval model that integrates fine-grained relevance signals distilled from an LLM directly into the retrieval function. 
Experiments on MS~MARCO show that FGR-ColBERT (110M) achieves a token-level F1 of 64.5, exceeding the 62.8 of Gemma~2 (27B), despite being approximately $245\times$ smaller. At the same time, it preserves retrieval effectiveness (99\,\% relative Recall@50) and remains efficient, incurring only a 
$\sim\!1.12\times$ latency overhead compared to the original ColBERT.

\keywords{Late Interaction  \and Token-Level Relevance \and LLMs.}
\end{abstract}

\vspace{-3.5em}

\begin{figure}[h!]
    \centering
    \begin{subfigure}{0.36\linewidth}
        \centering
        \includegraphics[width=\linewidth]{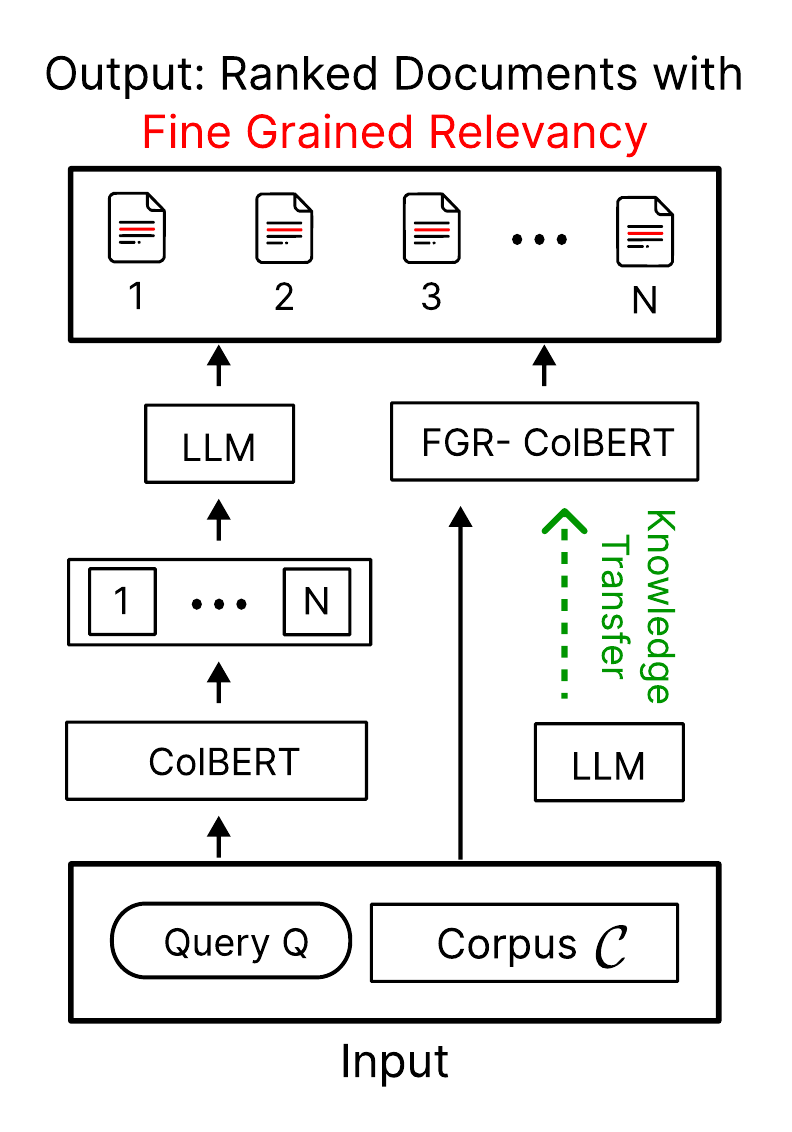}
        \caption{}
        \label{fig:problem-solution}
    \end{subfigure}
    \begin{subfigure}{0.60\linewidth}
        \centering
        \includegraphics[width=\linewidth]{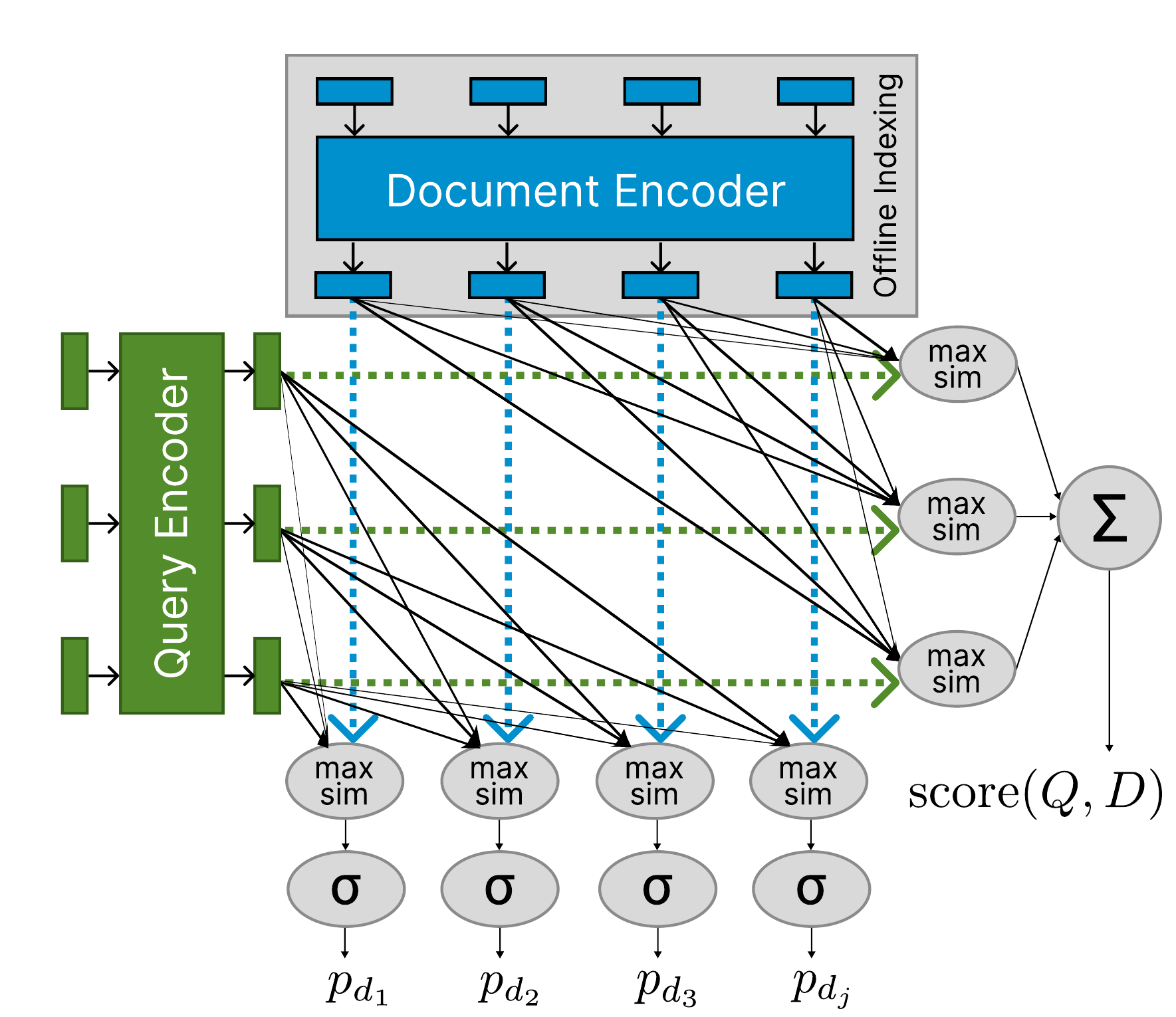}
        \caption{}
        \label{fig:interaction}
    \end{subfigure}
    \caption{(a) ColBERT retrieval followed by LLM span extraction vs.\ our approach with integrated LLM knowledge transfer. (b) Newly proposed late interaction with an added token-level relevance scoring, preserving document-level relevance.}
    \label{fig:combined}
\end{figure}

\section{Introduction}
Multi-vector dense retrieval models such as ColBERT~\cite{khattab2020colbert} achieve strong retrieval performance through efficient bi-encoder architectures~\cite{santhanam2022colbertv2}. However, beyond retrieving relevant documents, users often require precise evidence spans that directly answer a query~\cite{hashavit2024impact}. While large language models (LLMs) can identify such spans given a query–document pair, this typically incurs additional latency. Ideally, fine-grained relevance signals should be produced during retrieval itself, without requiring a separate post-processing step.

To address this, we propose \textbf{FGR-ColBERT}, a modification of ColBERT producing Fine-Grained Relevance signals directly during retrieval (cf. Figure~\ref{fig:problem-solution}). Since human span relevance annotation is costly, we use LLMs to augment the MS~MARCO~\cite{msmarco} dataset with evidence spans (e.g., containing answer). We then jointly train FGR-ColBERT for document-level retrieval via distillation from a cross-encoder and for token-level relevance prediction using LLM supervision.

Specifically, our contributions are as follows:%
\begin{description}[style=unboxed,leftmargin=0em,listparindent=\parindent,
    topsep=0pt,
    partopsep=0pt,
    itemsep=0pt]
    \setlength\parskip{0em}
    \item First, \textbf{we integrate fine-grained relevance signals into ColBERT.}
    We extend the ColBERT architecture to produce token-level relevance signals during retrieval by utilizing fine-grained supervision from LLM-generated evidence spans.
    
    \item 
    Second, we show the \textbf{proposed FGR-ColBERT achieves strong plausibility}—measured as token-level F1 agreement with human-annotated evidence spans—reaching 64.5 and matching or exceeding Gemma~2 (62.8), despite being $\sim\!245\times$ smaller (110M vs. 27B). We obtain this by distilling relevance signals from the 27B Gemma~2 model~\cite{team2024gemma}.

    \item 
    Third, \textbf{our approach maintains high retrieval effectiveness}, achieving recall@50 of 97.12 on a subset of MS~MARCO compared to the 98 baseline of the original ColBERT. Additionally, it has minimal overhead ($1.12\times$ latency) and no increase in index size.

\end{description}

\section{Method}


Let $\mathcal{Q}$ denote the set of possible queries and $\mathcal{D}$ the set of all documents. 
Given a query $Q \in \mathcal{Q}$ and a corpus $\mathcal{C} = \{D_1, \dots, D_m\} \subset \mathcal{D}$, 
the goal of document retrieval is to rank documents $D\in\mathcal{C}$ according to their relevance to $Q$.
Formally, the task is to learn a scoring function
\begin{equation}
    \text{score}(\cdot, \cdot) : \mathcal{Q} \times \mathcal{D} \rightarrow \mathbb{R},
\end{equation}
which assigns a relevance score to each query--document pair. Finally, documents are ranked in descending order of their scores, with higher-scoring documents considered more relevant to the query.

In this work, we consider the task of identifying fine-grained evidence spans within documents. 
For a given document $D$, let $\mathcal{S}(D)$ denote the set of all possible spans in $D$. 
We define a selection function
\begin{equation}
    \text{select}(\cdot, \cdot) : \mathcal{Q} \times \mathcal{D} \rightarrow \mathcal{P}(\mathcal{S}(D)),
\end{equation}
which returns a subset of spans in $D$ that are relevant to query $Q$.

\subsection{FGR-ColBERT}
Our method builds on the ColBERT~\cite{santhanam2022colbertv2} retrieval model, where each document $D$ and query $Q$ is encoded into set of contextualized token embeddings. Then, 
the relevance between a query $Q$ and a document $D$ is computed using a late interaction scoring function $\text{score}(Q, D)$, which aggregates token-level similarities by \textbf{matching each query token to its most relevant document token}:
\begin{equation}
\text{score}(Q, D) = \sum_{i=1}^{|Q|} \max_{j=1}^{|D|} E_{q_i}^\top E_{d_j}
\end{equation}
where $q_i$ and $d_j$ denote the $i^{th}$ and $j^{th}$ tokens in the query Q and document D, while $E_{q_i}$ and $E_{d_j}$ denote L2-normalized $h$-dimensional contextualized embedding vectors, respectively.

The core idea of our approach is to utilize ColBERT’s interaction mechanism orthogonally, i.e., \textbf{to find the best matching query token for each document token}, as illustrated in Figure~\ref{fig:interaction}. 
Applying an element-wise sigmoid activation yields a relevance probability $p_{d_i}$ for each document token, estimating the likelihood that a given token is relevant:
\begin{equation}
p_{d_i} \coloneqq \sigma \left( \max_{j=1}^{|D|} \hat {E}_{q_i}^\top \hat {E}_{d_j}\right).
\end{equation}
By thresholding the estimated probability, this formulation directly implements the selection function $\text{select}(Q, D)$.

We transform both query and document embeddings using a feed-forward network with a residual connection (omitting bias terms for simplicity)
\begin{equation} \label{eq:transform}
\hat {E}_{(\cdot)} \coloneqq E_{(\cdot)} + \mathrm{ReLU}(E_{(\cdot)} W_1) W_2,
\end{equation}
and weights $W_1\in \mathbb{R}^{h \times h_2}$ and $W_2 \in \mathbb{R}^{h_2 \times h}$.

\begin{description}[style=unboxed,leftmargin=0em,listparindent=\parindent]
    \setlength\parskip{0.3em}
    \item[Training.]
    We combine standard KL divergence for distillation from a cross-encoder reranker used in Colbert-V2~\cite{santhanam2022colbertv2} with a token-level binary cross-entropy objective for fine-grained supervision. The overall loss is defined as
\begin{equation}
\mathcal{L} = \mathcal{L}_{\mathrm{KL}} + \lambda \, \mathcal{L}_{\mathrm{BCE}}.
\end{equation}
Here, $\lambda$ is a weighting hyper-parameter and the binary cross-entropy term is given by
\begin{equation}
\mathcal{L}_{\mathrm{BCE}} = \frac{1}{|D|} \sum_{i=1}^{|D|} \left[ -t_{d_i} \log ( p _{d_i}) - (1 - t_{d_i}) \log (1 - p _{d_i}) \right],
\end{equation}
where $t_{d_i}\in\{{0,1\}}$ denotes ground-truth token-level relevance signals obtained from an LLM. Note that, the supervision of fine-grained relevance is intentionally applied exclusively to positive samples.
As a result, the model develops a bias toward always trying to identify a relevant span, since the $\mathcal{L}_{\mathrm{BCE}}$ function is never exposed to examples lacking a positive target.
This behaviour is not considered detrimental, as it encourages the model to consistently attempt to identify a relevant tokens within each passage.

\end{description}

\subsection{Enriching Dataset with Relevance Cues using LLMs}
LLMs, prompted with carefully designed instructions, serve as a direct implementation of the selection function $\text{select}(Q, D)$. For each query–document pair $(Q, D)$, the function returns a set of spans in $D$ that are relevant to $Q$, providing supervision signals for training.

\section{Experimental Setup}

Using this approach, we employ Gemma~2\footnote{We select Gemma~2 as the annotating model based on empirical observations indicating a strong alignment between its predicted evidence spans and human annotations, compared to alternative large language models.} to annotate both the training and development splits, resulting in the datasets \textit{MS-MARCO-Gemma-Train} and \textit{MS-MARCO-Gemma-Dev}, respectively.
To obtain human supervision, we additionally sample 140 query–document pairs from the MS~MARCO dev set  and task three annotators to label the relevant spans in each document, thereby obtaining annotations corresponding to $\text{select}(Q, D)$.

To evaluate the agreement between human ground-truth labels and model predictions, we compute the average token-level F1 score, which we refer to as \emph{plausibility}. Retrieval performance is assessed using Recall@k, measuring the proportion of queries for which at least one relevant document is retrieved within the top-$k$ results.

\section{Results}

\begin{table}[t]
\centering
\begin{minipage}[t]{0.57\linewidth}
\centering
\small
\begin{tabular}{lcccc}
\toprule
Model & Params & Hum. F1 & LLM F1 & R@50 \\
\midrule
FGR-ColBERT & 140M & 64.51 &  70.38  & 97.12 \\
ColBERT     & 140M & 51.67 &  50.08  & 98 \\
Gemma~2     & 27B  & 62.82 & -- & -- \\
\bottomrule
\end{tabular}
\vspace{0.3em}
\captionof{table}{Comparing parameter count, plausibility on human and LLM annotated dataset and retrieval (Recall@50) performance.}
\label{tab: main-results}
\end{minipage}
\hfill
\begin{minipage}[t]{0.4\linewidth}
\centering
\small
\begin{tabular}{lc}
\toprule
 & FFN \\
\midrule
Index increase & 0 \\
FLOPs & $4 n h h_{2} + n h_{2} + n h $\\
Time (ms)   & $0.7679 \pm 0.02$ \\
\bottomrule
\end{tabular}
\vspace{0.3em}
\captionof{table}{Resource overhead of the FFN architecture.}
\label{tab:resources}
\end{minipage}
\end{table}

\begin{description}[style=unboxed,leftmargin=0em,listparindent=\parindent]
    \setlength\parskip{0em}
    \item[FGR-ColBERT matches relevance cues obtained from LLM Gemma~2.]
    Table~\ref{tab: main-results} shows that the LLM achieves a token-level F1 of 62.82 on \textit{MS-MARCO-Gemma-Human}, which can be viewed as an approximate upper bound, as this LLM is used to obtain supervision. FGR-ColBERT slightly exceeds this value, reaching 64.51, although the difference may not be statistically significant. \emph{Despite having approximately 245$\times$ fewer parameters}, our FGR-ColBERT (110M) produces relevance cues during retrieval that are comparable to those of the much larger 27B model. 
    On the \textit{MS-MARCO-Gemma-Human} set, the model achieves even higher plausibility scores, further indicating successful knowledge transfer. Qualitative examples illustrating this alignment are provided in Appendix~\ref{sec:Examples}.

    \item[FGR-ColBERT retains 99\,\% of the retrieval performance.]
    Table~\ref{tab: main-results} presents retrieval results on the \textit{MS-MARCO-Gemma-Dev} set. The baseline ColBERT achieves a Recall@50 of 98, while FGR-ColBERT retains strong retrieval performance after training for fine-grained relevance estimation, reaching 97.1 (99\,\% relative).
    
\item[FGR-ColBERT incurs only a $\sim\!1.12\times$ computational overhead and does not increase index size.] 
Table~\ref{tab:resources} reports the additional resources required for fine-grained relevance extraction. Since the embeddings are transformed on-the-fly\footnote{Alternatively, storing precomputed transformed embeddings would lead to a $2\times$ increase in index size.} using a fully connected network (cf. Equation~\ref{eq:transform}) applied to the retrieved representations, no additional embeddings need to be stored. 

Let the embedding matrix $E_{(\cdot)} \in \mathbb{R}^{h \times n}$ and $h_2$ denote the hidden dimension of the network. We report the theoretical number of FLOPs required for the transformation. In practice\footnote{Measured on an NVIDIA GeForce RTX 3090 with $h=128$ and $h_2=768$.}, this corresponds to an average latency of 0.7679\,ms.
On the \textit{MS-MARCO-Gemma-Dev} set, retrieving the top-100 documents with the ColBERT framework requires $5.94 \pm 3.055$\,ms, resulting in an additional latency overhead of $1.13\times$.

\end{description}

\section{Conclusion}
We show that ColBERT can be extended to identify fine-grained relevance spans directly during retrieval via knowledge transfer from an LLM. Experiments on MS~MARCO demonstrate that FGR-ColBERT matches plausibility score of LLM Gemma~2 used for supervision while preserving retrieval effectiveness (99\,\% relative Recall@50) with only a $\sim\!1.12\times$ latency overhead.

Future work includes evaluating robustness on broader benchmarks such as BEIR~\cite{thakur2beir}, as we hypothesize that incorporating fine-grained signals distilled from LLMs can improve robustness.
Furthermore, we plan to extend the approach to long-document settings (e.g., LongEmbed~\cite{zhu2024longembed}), where fine-grained extraction is particularly beneficial.

\newpage
\bibliographystyle{splncs04}
\bibliography{bib}

\newpage

\appendix

\section{Theoretical Computation Time Increase of FGR-ColBERT }

\begin{table}
    \centering
    \begin{tabular}{l l c}
    \textbf{Computation Step} & \textbf{Expression} & \textbf{FLOPs} \\
    \midrule\midrule
    Up-Scaling & $A = E^R_d \cdot W_1$ & $2 n h h_2$ \\
    \cmidrule{1-3}
    Element-wise ReLU& $A' = \operatorname{ReLU}(A)$ & $n h_2$ \\
    \cmidrule{1-3}
    Down-Scaling & $B = A' \cdot W_2$ & $2 n h h_2$ \\
    \cmidrule{1-3}
    Residual Connection & $E^I_d = E^R_d + B$ & $n h$ \\
    \cmidrule{1-3}
    \textbf{Total FLOPs} & & $4 n h h_2 + n h_2 + n h$ \\
    \end{tabular}
    \caption{FLOP breakdown for the FFN network transforming embeddings for fine-grained relevance extraction.}
    \label{tab:ffn-computation-flops}
\end{table}

 The proposed architectural changes introduce computational overhead, as it applies two linear transformations with a non-linear activation in between, followed by a residual addition. For a document containing $n$ tokens, let $h$ denote the embedding dimension and $h_2$ the hidden dimension of the feed-forward layer. 

The computational cost can be derived using the standard approximation for matrix multiplication: multiplying an $a \times b$ matrix by a $b \times c$ matrix requires approximately $2abc$ FLOPs. 

\section{Qualitative Analysis of Fine-Grained Relevance Extraction }
\label{sec:Examples}

\begin{figure}
    \centering
    \includegraphics[width=0.95\linewidth]{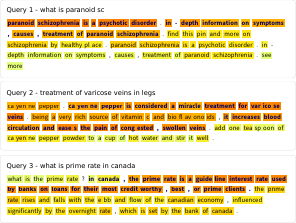}
    \caption{Three positive passage-query pairs and corresponding token-level scores (highlighted; darker is higher) derived from the FGR-ColBERT model and fine-grained relevance cues provided by Gemma~2 (bold).}
    \label{fig:qualitative-example-demonstration}
\end{figure}
Figure~\ref{fig:qualitative-example-demonstration} presents fine-grained extraction examples produced by the FGR-ColBERT model.

Each example is sampled from the MS~MARCO small development set, with scores obtained during model inference. In the first example, the query asks for a definition of paranoid schizophrenia. The model assigns higher scores to the beginning of the passage containing the definition, aligning well with the fine-grained LLM-based annotations (highlighted in bold). The remainder of the passage, while related to the topic, is less relevant to the query and receives appropriately lower scores. The final phrase, ``see more'', is irrelevant and correctly assigned the lowest score.

Similar score distributions and alignment with annotations are evident in the subsequent examples, confirming that the token-level cues provided by the model can serve as a plausible explanation

\end{document}